\documentclass[aps,prl,twocolumn,groupedaddress]{revtex4}

\usepackage{graphicx}

\def\rom{\everymath={\fam0 }\fam0 }	

\begin{document}


\title{Ultra-Low-Power Superconductor Logic}

\author{Quentin P. Herr}
\email[]{quentin.herr@ngc.com}
\author{Anna Y. Herr}
\author{Oliver T. Oberg}
\author{Alexander G. Ioannidis}
\affiliation{Northrop Grumman Systems Corp., Baltimore, MD 21240},


\date{\today}

\begin{abstract}
We have developed a new superconducting digital technology, Reciprocal
Quantum Logic, that uses AC power carried on a transmission line,
which also serves as a clock. Using simple experiments we have
demonstrated zero static power dissipation, thermally limited dynamic
power dissipation, high clock stability, high operating margins and
low BER. These features indicate that the technology is scalable to
far more complex circuits at a significant level of integration.  On
the system level, Reciprocal Quantum Logic combines the high speed and
low-power signal levels of Single-Flux-Quantum signals with the design
methodology of CMOS, including low static power dissipation, low
latency combinational logic, and efficient device count.
\end{abstract}

\pacs{}

\maketitle


Power consumption has increasingly become a limiting factor
in high performance digital circuits and systems. According to a
U.S. Environmental Protection Agency study \cite{EPA2007}, the demand
of servers and data centers in the U.S. is approaching 12\,GW,
equivalent to the output of 25 power plants. Here we show a new logic
family, Reciprocal Quantum Logic, that combines the low energy and
high clock rates of superconductor devices with the essential
qualities of CMOS, including low static power dissipation, low latency
combinational logic, and efficient device count. On a system level,
this yields a factor of 300 reduction in power compared to projected
nano-scale CMOS, while taking in account the overhead of cryogenic
operating temperature.

 Superconducting digital electronics has long been considered the
ultimate low energy alternative to CMOS \cite{ITRS04} based on the
fundamental advantages of lossless interconnect and fast, low energy
signal levels. Passive superconducting interconnect allows data
transmission without signal amplification. Superconducting
interconnects have a typical bandwidth of 700 GHz, which has allowed
serial data rates up to 60\,Gbs$^{-1}$ for chip-to-chip communication
\cite{herr02}.  Lossless interconnects would enable large systems with
high computational density, as compared to conventional systems where
interconnect dominates the total power budget.

Unlike transistor circuits, where dissipated power is set by device
size and materials, superconductor circuits are in the regime where
device size and power dissipation is set by the thermal noise
limit. The active device, the Josephson junction, generates quantum
accurate digital information in the form of Single Flux Quanta (SFQ)
of magnetic field $\Phi_0=h/2e= 2.06 \times 10^{-15}$\,Wb. Using
equivalent units $\Phi_0\approx2$\,mVps$=$2\,mApH illustrates that the
SFQ can exist as a transient voltage pulse across the Josephson
junction $V=\int_0^{\tau}Vdt\equiv \Phi_0$ or as a persistent current
in a superconducting inductive loop. For a typical minimum critical
current of 0.1\,mA at liquid helium temperature, the SFQ pulse energy
$E_{SFQ}=\int_0^{\tau}IVdt\approx I_c\Phi_0$ is only $1 \times
10^{-19}$\,J. This is only about three orders of magnitude above the
fundamental thermal Boltzmann limit, $k_BT$, and is the practical
limit for classical digital circuits operating with low bit error
rate. Beyond this limit there are only reversible computing and
quantum computing.

Numerous circuits with record-breaking clock rates have been
demonstrated in the Rapid-Single-Flux-Quantum (RSFQ) \cite{Likharev91}
logic family, including a static digital divider operating up to
770\,GHz \cite{Likharev99}, digital signal processors clocked at
20-40\,GHz \cite{Vernik07}, \cite{aherr08}, and a serial
microprocessor at 20\,GHz \cite{Tanaka06}. Cryocooler-mounted
prototypes have included a digital receiver used for satellite
communications \cite{Mukhanov08} and high-end router components
operating at 47\,Gbs$^{-1}$ port speed \cite{Hashimoto08}. However,
the technology suffers from high overhead in static power dissipation
and device count, which offset the energy advantages of SFQ data
encoding. RSFQ circuits use DC power delivered on a common voltage
rail via bias resistors, which is analogous to TTL logic and inferior
to CMOS due to static power dissipation. Ten times more power is
dissipated in the bias resistors than in the active devices even in a
fully active RSFQ circuit. While the power rail voltage is only about
1\,mV, it draws significant current, reaching 1\,A for a circuit with
10,000 Josephson junctions. This results in high parasitic heat load
in the cryopackage. Additional overhead is incurred in the timing
design; RSFQ uses an active clock distribution network, which leads to
significant accumulated jitter and timing variations based on device
parameters and data pattern statistics. Timing design of high speed
circuits results in a total device count that is dominated by the
clock distribution network \cite{Kataeva06}. RSFQ is pipelined on the
gate level, which enables high clock rates but also incurs high
latency. The new logic family described here circumvents each of these
limitations while preserving the fundamental property of SFQ data
encoding.

We report a new superconducting logic family, Reciprocal Quantum Logic
(RQL), that eliminates static power by replacing bias resistors with
inductive coupling to an AC transmission line that effectively powers
the devices in series and eliminates large ground return current. The
AC power also serves as a stable clock reference signal, preventing
accumulated clock jitter. The novel power supply is paired with a
novel data encoding. A logical ``one'' is encoded as a reciprocal pair
of SFQ pulses of opposite polarity.  During the positive half cycle,
the logic operation involves storage and routing of SFQ data
pulses. While the gates have internal state with respect to the
positive pulse, the trailing negative-polarity SFQ pulse serves as a
reset. This greatly simplifies gate design and produces combinational
logic behavior. Similar to CMOS, these combinational gates allow
multiple levels of logic per stage for low latency. Overall, RQL
combines the high speed and low-power signal levels of SFQ signals
with the design methodology of CMOS.

Using simple experiments involving logic gates and a 1600-device shift
register and logic gates, we have demonstrated that RQL is at once
high speed and low energy with a low bit-error rate. We measure energy
dissipation to be within a factor of 1000 of the thermal limit at
clock rates in the range 2-10\,GHz for the shift register, and
negligible BER of less than $10^{-40}$ for the logic gates while
maintaining operating margins of $\pm$30\%. AC power is supplied to
the circuit on superconductor microstrip transmission line, which also
serves as a passive clock distribution network. We show high stability
of the clock at frequencies up to 12\,GHz. The technology scales to
the one-million device level clocked at 6\,GHz with only a 6\,mW power
supply, amounting to only 15\,mA on a 50\,$\Omega$ line with dynamic
timing variation of only $\pm$1\% of the clock period. This indicates
that the technology is scalable to complex circuits at a significant
level of integration. Computational efficiency of the circuits is
nearly three orders of magnitude higher in terms of operations per
Joule compared to high performance CMOS. Taking into account that
superconductor circuits require a cryocooler, with efficiency of
1,000\,W/W achievable at 4.2\,Kelvin, RQL circuits offer a
system-level factor of 300 less wall-plug power dissipation. This
makes RQL technology attractive for many applications, including high
end computing.

\section*{Results}

{\bf Power dissipation.} RQL circuits have zero static power
dissipation, so for the first time dynamic power dissipation in a
superconducting SFQ circuit could be measured directly.  The clock
power is carried on 50\,$\Omega$ lines that return to room temperature
without termination on chip, allowing direct measurement of the
relative amplitude of the output waveforms for an inactive and fully
active circuit.  Because dynamic power dissipation is so small, the
experiment requires a circuit with a large number of Josephson
junctions and a low AC power amplitude with relatively high coupling
to the clock line. The shift register was chosen as a
convenient test vehicle.

\begin{figure}
\includegraphics[width=3.3in]{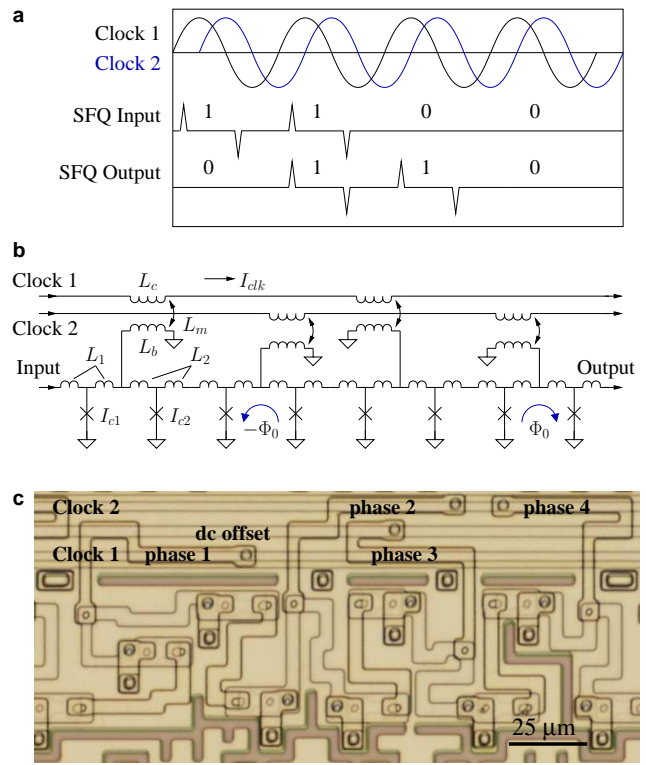}
\caption{ {\bf A Reciprocal Quantum Logic (RQL) shift register. a,}
Novel reciprocal data encoding, where the AC clock propagates digital
ones as pairs of Single Flux Quantum (SFQ) pulses of opposite
polarity. {\bf b,} Schematic of the RQL shift register
bit. Cross-wired transformers effectively produce a four-phase clock
with only two AC power lines, in quadrature. The SFQ pulses are shown
as loop currents that move through the circuit with a half cycle of
separation. Circuit parameters are: $I_{c1}=0.141$\,mA,
$I_{c2}=0.200$\,mA, $L_1=3$\,pH, $L_2=2.1$\,pH, $L_b=10.9$\,pH,
$L_m=5.1$\,pH, and $I_{clk}$=0.7\,mA amplitude and 0.2\,mA effective
offset (not shown). {\bf c,} Physical layout of the shift register in
a fabrication process with four Nb metal layers, with the middle
layers used for inductive wireup, and the top and bottom layers
serving as ground plane shields. The AC clock lines, and a separate
line to apply dc offset, are microstrips with the filament in a first
metal layer and ground in the top layer. Bias inductors lie in the
third metal layer situated on top of the clock signal line with strong
inductive coupling scaling linearly with the length of the
transformer.
\label{fig1}}
\end{figure}

Fig.~\ref{fig1} shows the schematic and physical layout of one bit
of the RQL shift register circuit. The four-phase clock is a fundamental
feature that provides directionality. Without this, the positive pulse
that moves forward during the positive half clock cycle would travel
backward during the negative half of the cycle, annihilating the
negative pulse. Instead, the positive pulse rides the leading edge of
the clock from one phase to the next and arrives at the output after
one cycle of delay, and the negative pulse follows with half a cycle
of separation.

A 200-bit shift register with 1600 Josephson junctions was fabricated
in a commercial superconductor fabrication process with 4.5\,kA\,cm$^{-2}$
Josephson junction critical current density, 1.5\,$\mu$m minimum
feature size, and four metal layers \cite{Hypres}. The junction plasma
frequency is 250\,GHz, corresponding to an SFQ pulse width of 3\,ps at
critical damping.  The impedance of the power transmission line,
implemented as microstrip with 2.3\,$\mu$m minimum line width and
SiO$_2$ dielectric thickness of 850\,nm, is limited to 32\,$\Omega$
including parasitic capacitive coupling to the bias inductors of about
7\,fF. Impedance matching to 50\,$\Omega$ was accomplished with
tapered microstrip lines leading to the contact pads.

\begin{figure}
\includegraphics[width=3.3in]{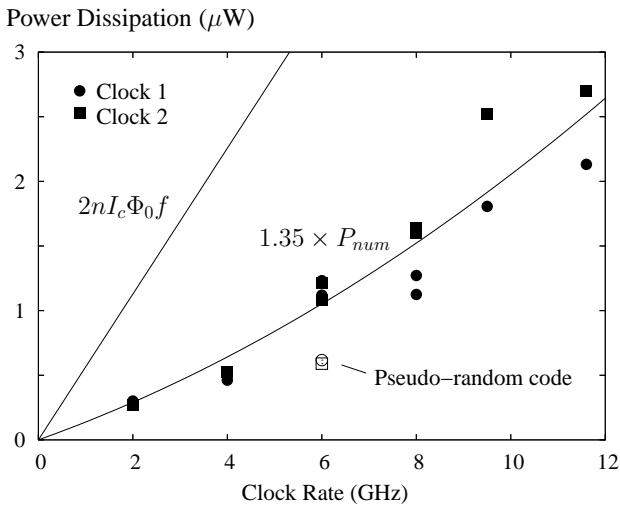}
\caption{ {\bf Power dissipation.} Power ratio of the clock output for
the two data patterns, corresponding to all ``ones'' and all
``zeros,'' was measured for frequencies of 2-12\,GHz. Power
dissipation is derived from the directly-observed power ratio and the
on-chip clock power of 12.5\,$\mu$W, calculated as the geometric mean
of applied and returned power.  At 6\,GHz and below, measurement error
is within the size of the data points. At 8\,GHz and above, the
primary source of error is variation in clock attenuation on the
different lines in the American Cryoprobe BCP-2 chip holder, producing
visible spread between data points for clock 1 and clock 2. Additional
data points correspond to a 6\,Gbs$^{-1}$ pseudo-random input pattern
that shows half the power dissipation compared to all ``ones'', as
expected. Measured power dissipation agrees well with the result from
circuit simulation, $P_{num}$, with a single multiplicative fitting
parameter. However, the power is three times smaller than the
analytical estimate that scales with circuit size and frequency as
$2nI_c\Phi_0f$ and that would apply to dc-powered SFQ devices.
\label{fig2}}
\end{figure}

The data in Fig. \ref{fig2} correspond to the power dissipated in a
fully active circuit with an all ``ones'' input pattern relative to
the inactive case of all ``zeros''. Using the estimate $2I_c\Phi_0$,
energy dissipation per digital ``one'' is $6.8 \times 10^{-19}$\,J
with average $I_c=170$\,$\mu$A. Total power dissipation of the circuit
would scale with the number of Josephson junctions and clock
frequency. Measured power dissipation in the circuit is three times
smaller than this estimate.  Additional data points corresponding to a
6\,Gbs$^{-1}$ pseudo-random input pattern show $0.6\,\mu$W total power
dissipation in the 800 Josephson junctions on each clock line. This is
half the power dissipation of the all ``ones,'' as expected, and is
only three orders of magnitude above the von Neumann-Landauer thermal
limit\cite{Landauer}, $k_BT\ln 2$ per bit.

A model for SFQ dissipation based on the energy potential indicates
that the work done on a switching junction is a function of bias
current rather than critical current \cite{BaronePaterno}.
Physical-level simulation \cite{whiteley} of the circuit shows that in
the range of interest, where clock period is much longer than the
switching time of the Josephson junctions, data pulses pass through
each stage early in the clock cycle under low bias conditions. This
results in a low energy SFQ pulse with simulated dissipation of about
$0.25I_c\Phi_0$ at 6\,Gbs$^{-1}$. Switching of the Josephson junctions
is shifted to higher bias at higher frequency, producing a slightly
non-linear frequency dependence. Experimental data agreess with this
simulation result fit with a single prefactor of 1.35.

{\bf Clock phase stability.} Switching of Josephson junctions not only
attenuates the AC clock but also adds accumulative delay.  The
magnitude of this effect can be estimated using a simple linear model
where the Josephson junction acts as an inductor if superconducting,
and as a resistor if switching. The clock propagation time in the case
of all digital ``ones'' is the same as for an isolated line
$\tau=\sqrt{L_cC_c}$=7.6\,fs/$\mu$m, where $L_c$ = 0.3\,pH/$\mu$m,
$C_c$ = 0.29\,fF/$\mu$m are the clock line inductance and capacitance
in the circuit. In the case of all digital ``zeros'', propagation time
is $\tau'=\sqrt{L'_cC_c}$ with inductance $L_c'$ given by the
impedance transformation for inductive coupling
\begin{equation}
L'_c = (1-k^2)L_c+k^2L_c(1||(L_g/L_b)),
\label{transf}
\end{equation}
where $L_b$ is the bias inductor and $k=L_m/\sqrt{L_cL_b}$ is the
magnetic coupling constant as shown in Fig. \ref{fig1}b. $L_g$ is the
inductance of the RQL gate connected to the bias inductor. In the
shift register, $L_g=(L_{J1}+L_1)||(L_{J2}+L_2)$ is the series and
parallel combination of the interconnect and the Josephson inductances
$L_J=\Phi_0/2\pi I_c$. In general terms, data-dependent phase delay
of the clock scales as $k^2$ and can be minimized by reducing coupling
to the clock line and increasing AC clock power.

Accumulated variable clock delay for the entire 200-bit shift register is
1.4$\pm$0.2\,ps and is independent of frequency.  Variable delay was
directly observed on the clock return from the chip at 2-12\,GHz on a
sampling oscilloscope to compare the data patterns of all ``ones'' and
all ``zeros''. Accuracy was limited by drift between the two
phase-locked synthesizers that clocked the chip and triggered the
oscilloscope. This result is in agreement with the analytical estimate
of Eq.~\ref{transf}, which gives 1.5\,ps variable clock delay for the
complete 200-bit shift register circuit.

\begin{figure}
\includegraphics[width=3.3in]{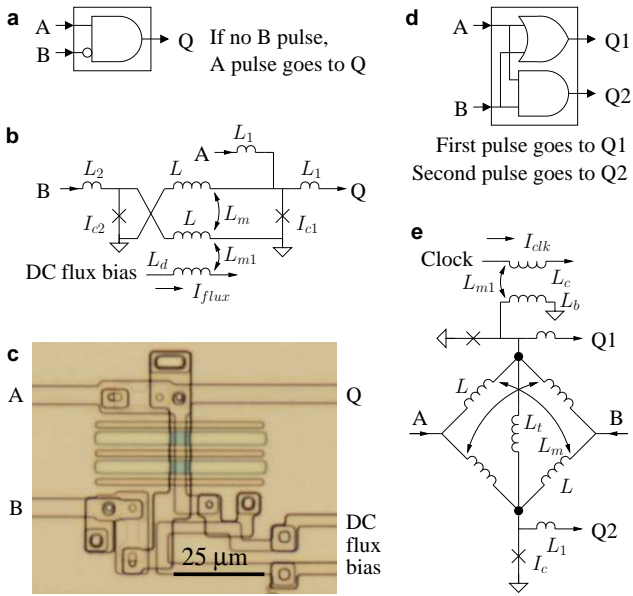}
\caption{ {\bf RQL logic gates.} The logic gates AnotB and AndOr are
simple and robust due to the reset function implicit in reciprocal
data encoding. {\bf a,b,} Block diagram, pulse logic behavior, and
schematic of AnotB gate. Operation of the gate is based on a
high-efficiency transformer that is cross-wired to invert the polarity
of the signal from input B.  Schematic values are $I_{c1}=0.141$\,mA,
$I_{c2}=0.100$\,mA, $L_1=3$\,pH, $L_2=2.1$\,pH, $L=20$\,pH,
$L_m=17$\,pH, $L_{m1}=2.5$\,pH, and $I_{flux}=0.4$\,mA.  {\bf c,} In
the physical layout, the high efficiency transformer is implemented
using moats in the upper and lower ground plane, which appear as seven
horizontal traces cutting across the vertical wires. {\bf d,e,} Block
diagram, pulse logic description, and schematic of AndOr gate. Here
the high-efficiency transformers aid propagation of either input pulse
to the outputs, but inhibit propagation to the opposite input.
Schematic values are $I_c=0.118$\,pH, $L=3$\,pH, $L_1=20$\,pH,
$L_m=16$\,pH, $L_t=20$\,pH, $L_b=12.8$\,pH, $L_{m1}=2.5$\,pH, and
$I_{clk}=0.7$\,mA amplitude and 0.2\,mA offset.
\label{fig3}}
\end{figure}

{\bf Logic gates.} Routing and processing of pulse-based signals is
distinct from transistor-based voltage-state logic, as shown in
Fig.~\ref{fig3}. Logical A-and-not-B (AnotB) means that an input pulse
A will propagate to output Q unless a pulse on input B comes
first. Logical And \& Or (AndOr) means that the first input pulse, if
any, goes to Q1 (logical OR), and the second input pulse goes to Q2
(logical AND).  Inputs to the AnotB gate must satisfy the timing
requirement that B arrive before A generates an output. There is no
similar timing requirement for the AndOr gate.

The logical behavior of the gates is based on the reciprocal data
encoding. Considering only the positive pulses, the gates are similar
to the state machines of RSFQ logic, as input changes the internal
flux state of the inductive loops. However, the trailing negative
pulse erases the internal state every clock cycle and produces
combinational logic behavior. The reset operation afforded by the
negative pulse greatly simplifies the logic design, so that each gate
consists of only two active devices with inductive interconnect. The
gates have large parametric operating margins, including simulated
tolerances on junction critical currents of at least {$\pm$50\%}.

\begin{figure}
\includegraphics[width=3.3in]{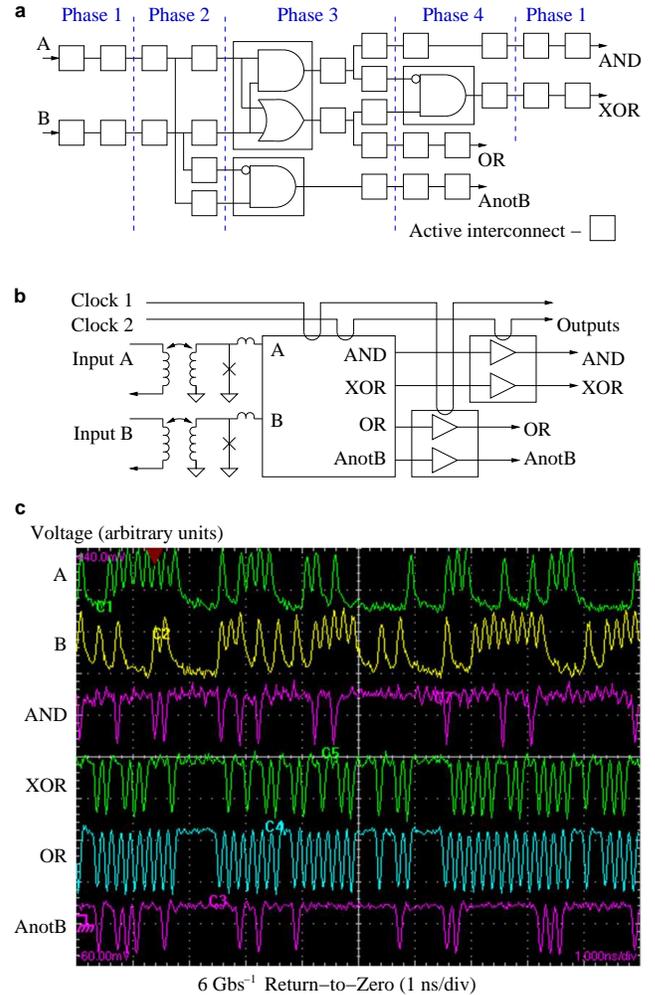}
\caption{ {\bf RQL logic test. a,} The circuit has two inputs,
three logic gates, and four logic outputs including a synthesized
XOR. Fanout is produced using interconnect consisting of two-junction
shift register stages. {\bf b,} The complete circuit includes input
gates that convert a return-to-zero (RZ) waveform to RQL data encoding
and on-chip output amplifiers that convert signals back to RZ voltage
levels.  The clock and data inputs are inductively coupled to the
circuit and return on another set of signal lines without contacting
chip ground, which contributes to high signal integrity at GHz
rates. {\bf c,} Input and output waveforms at 6\,Gb/s were captured on a
sampling oscilloscope.  A 1023-bit pseudo-random pattern was split and
applied to the inputs with a 39 bit offset. No signal averaging,
smoothing, or subtraction was used in the measurement.
\label{fig4}}
\end{figure}

\begin{figure}
\includegraphics[width=3.3in]{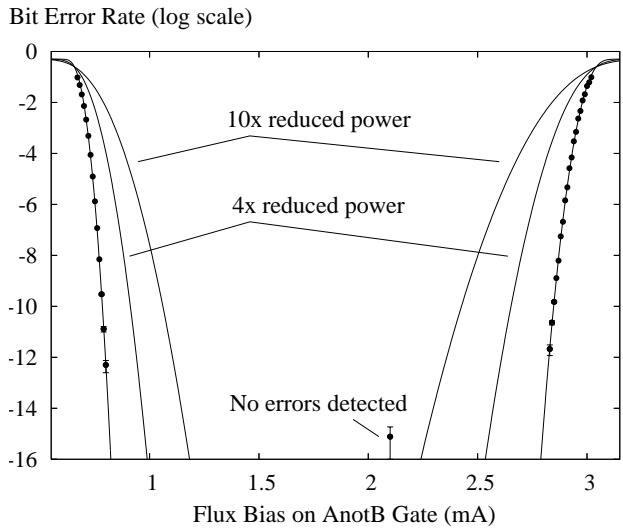}
\caption{ {\bf Bit-error rate (BER).} The BER of the AnotB gate at
6\,GHz is shown as function of its flux bias $I_{flux}$. A 32-bit
input pattern generated with an Anritsu MP1763C was split and applied
to the inputs with a 15 bit relative shift, and the XOR output was
compared to the correct pattern with an Anritsu MP1764C error
detector. Error bars on the lowest points correspond to counting
statics of 4 errors (left) and 5 errors (right). Near the center, no
errors detected for a period of 30 hours gives an error floor below
$10^{-15}$ for the entire circuit. The data fit to the error function
extrapolate to a minimum BER of $10^{-480}$ at the optimal bias of
1.82\,mA. Additional curves correspond to the BER scaled for reduced
device size and power.
\label{fig5}}
\end{figure}

On the physical level, both RQL gates have a bistable internal flux
state corresponding to $\pm\Phi_0/2$. The AnotB gate has an explicit
flux bias line that sets up positive current through both junctions to
ground. When either junction is triggered by an SFQ pulse, the flux
state is reversed, which reverses current through the junctions and
inhibits triggering of the opposite junction. The AndOr gate has an
implicit flux bias on inductor $L_t$, as the bias inductor is
connected at the end that favors the OR output. Triggering of OR
output redirects the bias current through $L_t$ to the AND output.
The gates require high efficiency transformers that provide high
common-mode inductance, when currents through the inductors are in the
same direction, and low differential mode inductance. In the AnotB
gate, the transformer inverts the polarity of the signal between input
B and the output. In the AndOr gate, the transformers aid propagation
of either input pulse to the outputs, but inhibit propagation of input
to the opposite input.

RQL logic gates cannot drive each other directly, but need at least
one interconnect cell to achieve fanout of one. The interconnect cell
is similar to the two-junction shift register unit cell, but with some
reduction of bias current to junctions interior to the clock phase
boundaries. The standard requirement of fanout equal to four can be
achieved using seven interconnect cells arranged in a binary tree. The
four phase clock provides an implicit pipeline without additional
devices for latches or clock distribution. Large circuit blocks with
multiple levels of logic can be on a single clock phase for low
latency, or smaller blocks can be used in a deeper pipeline for higher
maximum clock rate.  A crucial property of the RQL pipeline is
self-correcting timing. A pulse arriving early at a phase boundary
will be delayed due to low bias current. A late pulse will arrive
during high bias and will be accelerated. Pulses travelling through
the pipeline constantly tend toward timing equilibrium. However, at
too high a clock rate a late pulse will arrive at the phase boundary
on the falling edge of the clock and will be delayed further, causing
failure. The maximum clock frequency is determined by the number of
switching events per phase during approximately one quarter of the
clock period.

Fig.~\ref{fig4} shows the RQL logic test circuit. The circuit consists
of an AndOr gate, two AnotB gates, and interconnect.  The circuit has
two inputs and four outputs to produce the logic functions OR, XOR,
AND, and AnotB. Logical XOR is synthesized using an AndOr gate
followed by an AnotB. Total latency through the circuit is one clock
cycle. The timing requirement of each AnotB gate is satisfied by
placement on a clock phase boundary, with gate output controlled by
the trailing clock phase of the output interconnect. With this
arrangement the relative timing of inputs A and B is not important, as
both input are certain to arrive before the output is generated.

The logic circuit was collocated on the same chip as the shift
register. The complete circuit includes input gates that convert a
return-to-zero (RZ) waveform to RQL data encoding, and output
amplifiers that convert back to RZ voltage levels. The distributed
output amplifier, similar to that described in \cite{Herr10}, provides
a 2\,mV signal.  At the 1.5\,$\mu$m lithography node the logic circuit
itself was designed for 20\,GHz operation, but was limited in test to
6\,GHz by the input pattern generator. Measured operating margins on
clock power were $\pm$25\%, limited on the low end by the output
amplifers, and on the high end by overbiasing of the logic
circuit. Similar operating margins were measured at the lower frequency
of 2\,GHz.

{\bf Bit-Error Rate} SFQ circuits operate in the thermal limit, with
device size ultimately determining both energy dissipation and
bit-error rate (BER), which must be considered in conjunction with
each other.  Fig.~\ref{fig5} shows the BER of the AnotB gate clocked
at 6\,GHz, measured by monitoring the logical XOR output while setting
the flux bias on the gate to values near failure.  At the nominal
point, the flux bias provides symmetry between the bi-stable flux
states in the gate. At low bias, the observed failure is of logical
``ones'' becoming logical ``zeros'' uniformly throughout the pattern;
at high bias, logical ``zeros'' become logical ``ones''. Both error
modes indicate switching errors of the junction, labeled $I_{c1}$, that
generates the output.

The data are fit to a Gaussian distribution with bit-error rate
\begin{equation}
 p= \frac{1}{4}\rom{erfc} \left( \frac{\pm(I-I_t)/20}{\sqrt{2}\delta I}\right),
\label{erf}
\end{equation}
where $I-I_t$ is the distance of the flux bias current $I$ from the
error threshold $I_t$, and $\delta I$ is the root-mean-square noise
current. The prefactor takes into account that for either error mode
only half of the bits in the pattern contribute to the error rate. The
factor of twenty represents the transfer function between applied flux
bias and current induced in the loop containing the Josephson
junctions. Numerical fits were obtained with less than 1\% asymptotic
standard error. On the left, $I_t=0.66$\,mA and $\delta
I=1.02$\,$\mu$A; on the right, $I_t=3.04$\,mA and $\delta
I=1.56$\,$\mu$A. At the optimal bias point of 1.82\,mA, the data
extrapolate to negligible BER for the gate under test. The lowest rate
actually measured is below $10^{-15}$ for the entire circuit,
including the output data link.

\section*{Discussion}


The extrapolated minimum BER in our test indicates that device size
and power could be scaled down still further. The error mechanism may
involve either storage errors \cite{KleinMukherjee}, decision errors
\cite{Filippov91}, or even timing errors in an over-clocked circuit
\cite{LikharevRylyakov99}. In all of these cases, noise current scales
as the square root of the Josephson critical current, while the
current scale of the error threshold goes linearly.  Applying this
scaling to Eq.~\ref{erf} indicates that device size could be reduced
by a factor of ten and still extrapolate to a minimum error rate of
10$^{-44}$, which is negligible even for the most demanding
applications including high-end computing. As a practical matter, BER
below 10$^{-44}$ could be maintained over a wide flux bias margin of $\pm
30$\% by scaling device size down by a factor of four.  Measured noise
current in our test is consistent with previous results for the
gray zone of the RSFQ comparator \cite{Filippov97}, \cite{Filippov95},
but with much lower bit error rates relative to RSFQ circuits
\cite{HerrAPL96, Bunyk99, Herr99} due to the larger operating margins
in RQL.

The chip power scales linearly with number of junctions and
frequency. We measured that with a 12.6\,$\mu$W power supply, 800
junctions clocked at 6\,GHz have a worst-case data-dependent power
ratio of 0.91, corresponding to a variation in bias current amplitude
of $\pm 2$\%. Such a circuit scaled to $10^6$ junctions and with a
manageable maximum bias current variation of $\pm 10$\% would require
a 6\,mW power supply, amounting to only 15\,mA of current on a
50\,$\Omega$ line. On the same circuit we measured a 1.4\,ps
worst-case variable clock propagation delay, independent of frequency.
The given $10^6$ junction circuit would correspond to twenty times
less coupling to the clock line. Because variable clock delay scales
as $k^2$ this circuit would have a timing variation of only 5\,ps, or
only $\pm 1$\% of the clock period at 6\,GHz. The AC clock provides a
stable clock reference that suppresses accumulative clock jitter, so
ultimately clock frequency is limited by the switching time of the
Josephson junction, which scales linearly with feature size. At the
0.8\,$\mu$m lithography node, we can expect a 70\,GHz maximum clock
frequency \cite{Abelson04}, or alternately a 6\,GHz operation with
twelve levels of logic per pipeline stage.

AC power distribution on-chip will benefit from the exceptional
microwave properties of superconducting materials that have found
applications ranging from single-photon qubit resonators with Q of
$10^4$ to THz dark matter detectors \cite{Vion08,
Detectors}. Monolithic integration of RQL gates with microwave
components, including power splitters, matching networks, and phase
shifters, is a strength of the technology. Power dissipation in these
passive components is as low as 1\% per wavelength \cite{Nblosses,
Nblosses1THz}, which would correspond to only 2.3\% of the applied
power in our shift register experiment. In the cryopackage, a 15\,mA
amplitude for the clock is consistent with low heat transport on the
wires \cite{Cryo-Gupta}. Very high clock rates up to 71\,GHz have
already been demonstrated for Josephson voltage standards using
waveguides in the cryopackage \cite{PTB, NIST}.

Computational efficiency of the measured circuits are approaching
$1000\,k_BT$ with further reductions expected using smaller devices,
giving unmatched efficiency in terms of operations per Joule.  This
means the technology offers a low energy solution for high end
computing even after taking into account the overhead of the
cryocooler, on the order of 1000\,W/W at 4.2\,K \cite{Coolers}.
Because the 700\,GHz energy gap in Nb makes superconductors inherently
radiation-hard \cite{King91}, the technology may be useful for
computationally intensive applications in space.  Since device size
and power can be scaled with temperature to remain in the
noise-limited regime, the technology would be ideal for classical
control, readout, and error-correction feedback for solid state qubits
\cite{QubitControl} operating at millikelvin.

\begin{acknowledgments}
This work was supported in part by the Defense Micro-Electronics
Activity under the Advanced Technology Support Program. The authors
thank Marc Manheimer for discussion, John Fusco for administration,
and acknowledge assistance from Isaac Carruthers with the software
design environment, from Donald Miller and Steve Shauck with the
design.
\end{acknowledgments}

\end{document}